\documentclass{article}

\usepackage{amssymb}
\usepackage{graphicx,epstopdf}
\usepackage{authblk}
\usepackage{subfig}
\usepackage{lipsum}
\usepackage{amsfonts}
\usepackage{graphicx}
\usepackage{epstopdf}
\usepackage{amsmath}
\usepackage{algorithm}
\usepackage[noend]{algpseudocode}

\usepackage{enumitem}
\setlist[enumerate]{leftmargin=.5in}
\setlist[itemize]{leftmargin=.5in}

 \title{Vectorized Uncertainty Propagation and Input Probability Sensitivity Analysis}

 \author{Kevin Vanslette, Arwa Alanqari, Zeyad Al-awwad,\\ and Kamal Youcef-Toumi }
 \affil{\small{\emph{Massachusetts Institute of Technology}}}

\affil{ \small{\emph{ King Abdulaziz City for Science and Technology}}}
 \date{June 4, 2019}

\begin{document}
\maketitle

\begin{abstract}

In this article we construct a theoretical and computational process for assessing Input Probability Sensitivity Analysis (IPSA) using a Graphics Processing Unit (GPU) enabled technique called Vectorized Uncertainty Propagation (VUP). VUP propagates probability distributions through a parametric computational model in a way that's computational time complexity grows sublinearly in the number of distinct propagated input probability distributions.  VUP can therefore be used to efficiently implement IPSA, which estimates a model's probabilistic sensitivity to measurement and parametric uncertainty over each relevant measurement location. Theory and simulation illustrate the effectiveness of these methods.

\end{abstract}
%
\section{Introduction}
    %
Mathematical and computational models serve as convenient representations of physical and human-made processes. When data is difficult to obtain, unavailable (e.g., in the case of future value prediction), or computationally intractable, data-driven or physics based models, 
are designed to inexpensively and reliably reproduce, and extrapolate, data. In general, a model may be deterministic or nondeterministic depending on the nature of the physical or human-made process it is trying to represent.

A model, its inputs, model form, and outputs, may be uncertain for any number of reasons. A model may have uncertain model parameters or noisy input data values, uncertainty pertaining to computational representation (discretization or surrogate mode error), model form uncertainty (i.e. how do we know this is the best model?), and others. It is common in the literature to categorize such uncertainties as being either: \emph{aleatory} and naturally stochastic, or \emph{epistemic} and deriving from a lack of knowledge on part of the modeler \cite{Mullins}. 

Uncertainty may be propagated through a computational model using existing sampling methods. This includes simple Monte Carlo (MC), stratified sampling, and importance sampling, as well as simple quadrature integration such as midpoint, trapezoidal, or other integration rules. When the integrals are high dimensional or over complicated probability distributions, their results may be approximated using Markov Chain MC \cite{MCMC}, Metropolis Algorithm \cite{Metro}, Gibbs Algorithm \cite{GA}, Hamiltonian MC \cite{stan}, or slice sampling \cite{SS}.  Other effective methods for propagating uncertainty construct surrogate models such as truncated Polynomial Chaos Expansions (PCE) \cite{Spectral, 134, 153,NumChallenges}, Gaussian Process \cite{KOH} models, and discretized models \cite{18,62,63,14,20}, that are sampled instead of the original models, which introduces additional uncertainty that in principle can be propagated at the model verification stage \cite{Shankar,Li,Roy2}. 

The state of the art method for performing computational model sensitivity analysis (SA) involves constructing Variogram Analysis of Response Surfaces (VARS) \cite{VARS1, VARS2,GUPTA}. Integrating the VARS variogram leads to Integrated VARS (IVARS), which is able to provide a ``characterization of sensitivity across the full spectrum of scales" in a computationally efficient manner. IVARS includes variance based (Sobol) \cite{Sobol1,Sobol2,UQTk,Dakota} and derivative based \cite{Morris} sensitivity analysis methods as special cases. Each relevant scale is given equal weight in IVARS. 

Performing SA on the basis of local and global input measurement uncertainties across the set of possible measurements appears to be a highly relevant, yet omitted topic, from modern SA.\footnote{Perhaps due to its presumably high computational time complexity.} As the task of collecting high precision data may be costly, ultimately we would like to know how our lack of exact knowledge over the inputs of a model affects our knowledge of the output so we can focus our collection efforts in regions with the highest payoff. 

In the literature, the main application of GPU computing in computational probability theory is toward solving the steady state Markov chain problem \cite{MarkovChainGPU1,MarkovChainGPU2,MarkovChainGPU3} and toward speeding up probabilistic machine learning \cite{tf}, i.e., probabilistic model parameter learning. The steady state Markov chain problem involves finding the steady state probability distribution of a Markov chain system. 
GPU solutions to this problem involve representing an initial probability distribution as a vector and applying a probability valued transition matrix on it many times until an approximate steady state solution is found. 
Because GPU's are capable of multiplying matrices with high efficiency, \cite{MarkovChainGPU1,MarkovChainGPU2,MarkovChainGPU3} utilize the GPU for this process. When applicable, the sparseness of the matrix can be utilized to improve computational speed and memory requirements \cite{sparse}. A method in which many different input probabilities are sent through the GPU and propagated through a computational model ``simultaneously" has yet to be explored in the context of uncertainty quantification.



In this article we construct a theoretical and computational process for assessing Input Probability Sensitivity Analysis (IPSA) using a GPU enabled Vectorized Uncertainty Propagation (VUP) technique. We build IPSA from the theoretical foundations of VARS and IVARS and thus it can reproduce their results, and thus the results of other SA methods, as special cases. A full IPSA relies on an efficient method for propagating many input uncertainties through the computational model in question. By vectorizing the uncertainty propagation process and extending it to be able to evaluate many different input probability distributions ``simultaneously", 
we instantiate a VUP technique that is more computationally efficient at this task than MC. 
Thus, by using VUP for IPSA, IPSA becomes a rapid and efficient method for performing context specific SA. 

The main SA object generated by an IPSA is a set of output probability distributions that encode pointwise deviations from the estimated values of the model function as a whole, due to local input uncertainties.  These local output deviation probabilities can be used to calculate any number of quantities pertaining to local sensitivity such as expectation values, confidence intervals, maximum probability estimates, and others. The resulting distribution from marginalizing over the local distributions can be used to calculate global sensitivity quantities that incorporate the aggregate effect of uncertainty into the analysis.

IPSA differs from other probabilistic sensitivity analysis found in the literature. In IPSA, one uses a parametric model function (with uncertain parameters in general) directly rather than a data driven Gaussian Process \cite{PSA1}, which allows us to better address asymmetries in the model function output pdfs. 
Further, we are interested in the probabilistic response of the function over may different locations, rather than just one \cite{PSA2}, which gives us access to global and local measures of sensitivity and provides the connection to IVARS as a special case.






 The remainder of the article is as follows. We discuss computational models and uncertainty propagation in Section \ref{section2}. VUP is introduced in Section \ref{section3}. VUP and MC computational time complexity comparisons are derived and simulated in Section \ref{section4}. In Section \ref{section5} we derive IPSA from the foundational basis of IVARS and we illustrate an example IPSA problem using VUP for several input uncertainty scenarios. 



\section{Computational Models and Uncertainty Propagation\label{section2}}

A computational model may be represented mathematically as a function \cite{Shankar,Li,Roy2,UQTk,Dakota}.  Computational model functions may encode arbitrary input/output computer functions including but not limited to: differential and non-differential equations or their solutions, conditional piecewise functions, vector and matrix valued functions, or even computational functions of strings. The majority of the computational models above can be represented mathematically as a general function,
\begin{eqnarray}
M(\vec{v})=\vec{y},\label{modelfunction}
\end{eqnarray}
that maps inputs $\vec{v}\in\mathbb{R}^n$ to outputs $\vec{y}\in\mathbb{R}^{m}$.


We find it conceptually convenient to partition the input vector,
$$\vec{v}=(\vec{x},\vec{\alpha}),$$
into the vector $\vec{x}$ and $\vec{\alpha}$, which represent the incoming ``data like" inputs and ``model parameter like" inputs, respectively. The $\vec{x}$ inputs tend to vary from instance to instance and originate from outside of the system, whereas the $\vec{\alpha}$ inputs are internal to the model and are (in general, probabilistically) regressed to fit known data. 

In principle, computational models may be considered to be deterministic even if they have functional dependence on internally generated pseudorandom numbers. Because computers are deterministic machines, a computational model having internally generating pseudorandom numbers, $\vec{\lambda}$, according to a given pdf, $\rho(\vec{\lambda})$, is deterministic too. Thus, when appropriate, the $\vec{\lambda}$'s may be pulled out and lumped into the model parameters,
$$\vec{\alpha}\rightarrow\vec{\alpha}\equiv(\vec{\alpha},\vec{\lambda}),$$
without loss of generality.

Deterministic computational models can be represented using probability distributions that indicate complete certainty. 
A completely certain input value $\vec{v}$ that propagates through the deterministic model function $M$ to the output $\vec{y}$ with certainty is represented probabilistically as,
\begin{eqnarray}
\rho(\vec{y}|M,\vec{x},\vec{\alpha})=\delta(\vec{y}-M(\vec{x},\vec{\alpha})),\label{1}
\end{eqnarray}
where $\delta(...)$ is the Dirac delta function (or the indicator function $\Theta(...)$ in the discrete case).  We will call the left hand side of (\ref{1}) the ``model propagator", of which the right hand side is a ``deterministic model propagator". This equation represents the fact that we know computers are deterministic machines.

The goal of uncertainty propagation (UP) is to estimate the probability distribution function of the model outputs, $\rho(\vec{y}|M)$, due to a known amount of uncertainty in the input variables, $\rho(\vec{v}|M)$.  Theoretically, the resulting value of the output pdf from UP is given by marginalization over the uncertain inputs,
\begin{eqnarray}
\rho(\vec{y}|M) = \int_{\vec{x},\vec{\alpha}} \rho(\vec{y}|M, \vec{x},\vec{\alpha})\rho(\vec{x},\vec{\alpha}|M)\,d\vec{x}\,d\vec{\alpha},\label{UP}
\end{eqnarray}
which is an integral that must be estimated for each viable element in $\{\vec{y}\}$. In the case where all of the $\vec{\lambda}$ parameters are lumped into $\vec{\alpha}$, we may substitute equation (\ref{1}) into (\ref{UP}) and still represent models having random numbers within them. 

For the purpose of this paper, we will call any model where we do not know the functional form of the model propagator to be a ``black box model propagator" and for any model propagator that its functional form is known (e.g. (\ref{1}), Gaussian, etc.)  to be a ``computationally deterministic model propagator". From our point of view, one is dealing with a black box model propagator if, for one reason or another, it is impossible or computationally unrealistic to lump the internal (and potentially nested) random numbers $\vec{\lambda}$ into $\vec{\alpha}$. In such a case, MC type methods tend to bypass the need for constructing the functional form of the model propagator in favor of simply generating estimates of $\rho(\vec{y}|M)$. 

In this article we will only consider computationally deterministic model propagators. 
The models within this class are not limited to:  non-stochastic parametric models, models with known (or well estimated) functional forms of the model propagator (e.g. Markov Chain models, Gaussian Processes, etc.), or stochastic models with extractable low dimensional $\vec{\lambda}$ parameters. A given computationally deterministic model propagator may be simple or computationally complex to evaluate, e.g., they could just be algebraic functions or have internal search/optimization type routines. 

\section{Vectorization Uncertainty Propagation\label{section3}}
We begin vectorizing the UP process for computationally deterministic models by first representing these sets of integrals $\{\rho(y|M)\}$ as set of sums $\{p(y_i|M)\}$, where $p(y_i|M)$ is the probability of $y_i$ given the model $M$. Let $\overrightarrow{|\bigtriangleup|}_{y_i,M(j,k)}$ be the vector resulting from the componentwise modulus of the subtracted vectors, $\vec{y}_i-M(\vec{x}_j,\vec{\alpha}_k)$. Given that UP will be calculated computationally, we represent the problem in the discrete setting, $\rho\rightarrow p$,
\begin{eqnarray}
p(\vec{y}_i|M)&=&\sum_{j,k}\Theta\Big(\overrightarrow{|\bigtriangleup|}_{y_i,M(j,k)}\leq \vec{b}\Big)\,p(\vec{x}_j,\vec{\alpha}_k|M),\nonumber
\end{eqnarray} 
where $\Theta(B)$ is the discretized deterministic model propagator and the indicator function, which is equal to one if $B$ is true and is zero otherwise. The bin widths vector $\vec{b}$ uniformly partitions $\{\rho(\vec{y}|M)\}$ into $\{p(\vec{y}_i|M)\}$. 
We will notationally suppress the indices and the vector arrows and instead write,
\begin{eqnarray}
p(y|M)=\sum_{x,\alpha}\Theta\Big(|\bigtriangleup|_{y,M(x,\alpha)}\leq b\Big)\,p(x,\alpha|M),\label{4}
\end{eqnarray}
when there is no room for confusion. 

We vectorize the UP process by representing the discretized model propagator as a matrix and by performing matrix multiplication. Let $P(x,\alpha|M)$ represent an input probability vector, which has components equal to the discrete input probabilities $p(x,\alpha|M)$. The dimension of $P(x,\alpha|M)$ is equal to the number of samples in the joint input space $N=N_x*N_{\alpha}$, where $(N_x,N_{\alpha})$ is the number of samples per $(x,\alpha)$, respectively. Let the model propagator be represented by what we call the ``model matrix" $\mathcal{M}_{y, (x,\alpha)}$, which has components given by a model's discretized model propagator $p(y|M, x,\alpha)$ and which has dimension $N_y\times N$.  UP is performed by matrix multiplying the model matrix on the input probability vector. This generates $P(y|M)$, which is an $N_y$ dimensional output probability vector with components equal to the output probabilities $p(y|M)$. 
That is, UP from equation (\ref{4}) is,
\begin{eqnarray}
P(y|M)= \mathcal{M}_{y, (x,\alpha)}\cdot P(x,\alpha|M).\label{5}
\end{eqnarray}
In the computationally deterministic case $p(y|M, x,\alpha)=\Theta\Big(|\bigtriangleup|_{y,M(x,\alpha)}\leq b\Big)$, (\ref{5}) looks something like (where the $M$ is notationally suppressed),
\begin{eqnarray} \left( \begin{array}{c}
p(y_1)\\
\vdots\\
p(y_{N_y})\\
\end{array} \right)=\left( \begin{array}{ccccc}
1 & \dots&1 & \dots &0 \\
\vdots&\vdots&\vdots&\vdots &\vdots\\
0 & \dots&0&\dots & 1
\end{array} \right)\cdot
\left( \begin{array}{c}
p(x_1,\alpha_1)\\
\vdots\\
p(x_{j},\alpha_k)\\
\vdots\\
p(x_{N_x},\alpha_{N_{\alpha}})
\end{array} \right),\label{MUP}
\end{eqnarray}  
for a particular discretized deterministic model propagator, i.e., for a particular population of 1's and 0's in the model matrix. From the above equations, it becomes clear that the model propagator uses the model function to effectively sort and sum the input probabilities into their respective output probability bins. Equation (\ref{5}) is the vectorized representation of UP (i.e. VUP) we seek to utilize GPU computing. Due to normalization, the model matrix is usually (extremely) sparse, which can be taken advantage of to improve speed and memory requirements, that is, using \cite{sparse}.

The matrix representation of UP is purely for computational convenience. Standard matrix methods, such as applying a matrix inverse or pseudoinverse, can result in negative probabilities or other nonsensical results.

VUP has access to Bayes Theorem,
$$p(x,\alpha|y,M,\ell)=\frac{p(y|x,\alpha,M,\ell)p(x,\alpha|M,\ell)}{p(y|M,\ell)},$$
which is the probabilistically correct way to make inverse or backward type inferences. We represent this set of probability distributions $\{p(x,\alpha|y,M,\ell)\}\rightarrow \mathcal{M}_{(x,\alpha),y}\sim\mathcal{M}^{-1}_{y,(x,\alpha)}$ with what can be called a ``inverted model matrix" -- it maps output values to input values $y\rightarrow M^{-1}(y)= \{(x,\alpha)\}_y$.\footnote{Bayes theorem provides the rules for ``inverting" the model matrix as one can prove $\mathcal{M}_{(x,\alpha),y} \cdot\mathcal{M}_{y, (x,\alpha)}\cdot P(x,\alpha|M)= P(x,\alpha|M)$.}  
It should be stressed that this equation is \emph{predefined by the forward propagation} (i.e. equation (\ref{4})) and thus the inference is based purely on that instance (or context) represented by the input pdf $ P(x,\alpha|M)$. Because model functions $M$ are \emph{not} uniquely invertible in general, Bayes Theorem ends up assigning nonzero and nonunity probabilities over the possible inputs -- i.e., the multiple solutions of noninvertable functions are assigned probabilities.


We can use VUP to propagate a large number of distinct input probability distributions through a model matrix given the model matrix is an adequate representation of the model. Let each input probability distribution be distinct and differ by a given proposition or parameter $\ell$, of which there are $L$ many. Constructing an $N$ by $L$ input probability matrix, 
$$\mathcal{P}_{(x,\alpha),\ell}=\Big[ P(x,\alpha|M,\ell=1), ..., P(x,\alpha|M,\ell=L)\Big],$$
and operating the model matrix on it,
\begin{eqnarray}
\mathcal{P}_{y,\ell}= \mathcal{M}_{y, (x,\alpha)}\cdot \mathcal{P}_{ (x,\alpha),\ell},\label{VUP}
\end{eqnarray}
we propagate many distinct input probability distributions through the model ``simultaneously". Each column of $\mathcal{P}_{(x,\alpha),\ell}$ is an input probability vector that propagates to its corresponding output probability vector $ P(y|M,\ell)$, similarly situated in the output probability matrix $\mathcal{P}_{y,\ell}$. 

Due to the current trend in AI and machine learning, GPU's are expected to continue to improve at an accelerated rate and become more readily available, which makes VUP increasingly attractive. We expect that by further developing batching, smart or representative sampling, and distributed computing techniques, that we would be able to mitigate memory constraints in a way that would expand VUP's domain of applicability to higher dimensions over time. The input pdfs are discretized using the composite midpoint rule for rapidity. Given the recent results of \cite{quadMC}, which indicate that the composite midpoint integration rule is just as good an integral estimation technique as MC, we are optimistic about extending VUP to higher dimensions accurately.

\section{The Rapidity of VUP\label{section4}}

Appendix \ref{appendix} compares the computational complexity estimations of VUP and MC in the single and multiple propagated probability cases, which we will summarize here. These estimates comes from a few assumptions that can be relaxed if desired; however, similar results follow.

The estimation of the difference in the computational time complexity between MC ($C_{MC}$) and VUP ($C_{VUP}$) favors MC for the propagation of \emph{a single pdf} by an amount that is on the order of the number of samples,
\begin{eqnarray}
C_{MC}-C_{VUP}\sim -\mathcal{O}(N) \label{singleCC},
\end{eqnarray}
due to the sparseness of the model matrix multiplication (\ref{A5}). 

Because VUP reuses of the model matrix in (\ref{VUP}) for each propagated probability vector, the structure of the model matrix sorts the input probabilities automatically and therefore we do not need reevaluate the model function and resort the results as $L$ increases. Our method therefore increases sub-linearly in $L>1$, $C_{VUP}(L)<L*C_{VUP}$, whereas MC increases linearly, $C_{MC}(L)=L* C_{MC}$. The estimated difference in the computational time complexity for propagating $L$ input probability distributions is then, 
\begin{eqnarray}
C_{MC}(L)-C_{VUP}(L)\sim\mathcal{O}\Big((L-1)*N*(C_{model}+\log(N)+1)-L*N\Big),\label{multipleCC}
\end{eqnarray}
where $C_{model}$ is the computational time complexity of a model evaluation. Our method is favored for,
\begin{eqnarray}
\frac{1}{L-1}\lesssim\mathcal{O}\Big(C_{model}+\log(N)\,\Big),
\end{eqnarray}
which is almost always the case, given $L>1$ (\ref{A8}). Thus, we can expect large improvements to the computational time complexity when $C_{model}$, $N$, or $L$, is large. We will demonstrate these expected computational complexity trends experimentally in the next subsection.

\subsection{Testing the Rapidity of VUP\label{sectiontest}}
\begin{figure*}[h]
     \centering
            \includegraphics[width=.75\textwidth]{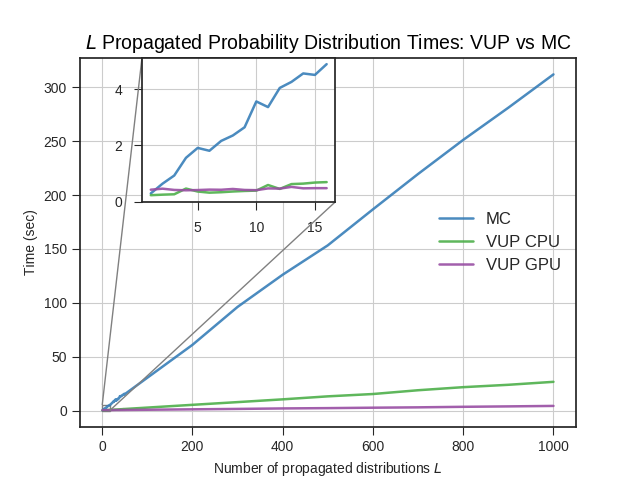}

            \caption[]
        {\small 
        We follow the experiment described in Section \ref{sectiontest} and find results to agree qualitatively with the computational complexity estimates in equations (\ref{singleCC}) and (\ref{multipleCC}). The figure plots clock times for propagating $L$ distinct probability distributions using a simple MC method, VUP on the CPU, and VUP on the GPU. The small $L$ behavior is depicted in the zoomed in window.
        The machine used for this test was leased through Paperspace web services and has the following specs: 8 CPUs, 30 GB of memory, and a Quadro P4000 (8 GB) GPU.

        } 
        \label{timeplots}

    \end{figure*}

Here we outline the time complexity simulation plotted in Figure \ref{timeplots}. Compared are the clock times for simple Monte Carlo using NumPy, VUP on the CPU using NumPy, and VUP on the GPU using PyTorch, for propagating $L$ distinguishable input probability distributions through a simple two dimensional computationally deterministic and vectorizable model function $y=M(x,\alpha)$.\footnote{The plotted trends are insensitive to small changes in the functional form of vectorizable model functions. We used $y=M(x,\alpha) = 1.1\sin{(x)} + 7 \sin^2{(\alpha)}$ in this simulation.} The number of samples per propagated distribution is constant and set to $N=10^6$ in this simulation. 

For the sake of this experiment, each input probability distributions is a two dimensional Gaussian distribution that has a single free parameter $\mu_{x}(\ell)$, giving a set of $L$ different input probabilities. Simple Monte Carlo follows Algorithm 9.1 and VUP on the CPU and GPU follow Algorithm 9.2 in Appendix \ref{appendix}. These algorithms are extended to many input probabilities. 

For the triple (MC, VUP CPU, VUP GPU), respectively, we find $L=1$ times of about\\ $(0.29,0.24,0.43)$ seconds and $L=1000$ times of about $(311.93,26.74,4.39)$ seconds. Thus, each method increases in time by a factor of approximately (1000, 100, 10), with respect to itself, as $L$ is increased from 1 to 1000. These times follow the expected trends from the computational time complexity analysis, (\ref{singleCC}) and (\ref{multipleCC}), respectively. Not included in the measured times for VUP GPU (and the computational complexity estimate) is the $\sim3$ second GPU initialization time. If the GPU initialization time is included for every instance, it acts as a 3 second time offset that causes the VUP CPU and VUP GPU lines to cross at around $L=90$ instead of at about $L=13$; however, the GPU only needs to be initialized once per session, not per instance, so this time was ignored.

Because MC has to evaluate the model $N*L$ times whereas VUP only has to evaluate the model $N$ times, small changes to the model function evaluation time complexity can cause large differences in the overall computation time. Because $L=1000$ and $N=10^6$ here, if the computational complexity required to evaluate the model function is increased by just a tenth of a millisecond, it leads to a 1.15 day increase in computation time for simple MC verses a 100 second computational time increase for VUP. Rather than being 70 times faster, as it was in our $L=1000$ experiment, the domination of the model function evaluation time complexity leads to VUP being about $L$ times faster than MC.

\section{Input Probability Sensitivity Analysis\label{section5}}
 
 In this section we develop Input Probability Sensitivity Analysis (IPSA), which naturally utilizes VUP. Because VUP can propagate many different probabilities through a model matrix rapidly, VUP allows us to access areas of UQ that may have previously been seen as inaccessible.  
 IPSA uses VARS and IVARS \cite{VARS1, VARS2} as a theoretical foundation and therefore we will review these SA methods for convenience.

 \subsection{SA, VARS, and IVARS Review}
 Although local SA is well defined in terms of local partial derivatives of the model outputs with respect to coordinates in the input space, the meaning and objectives of global SA remained somewhat unstructured until \cite{Gupta}. In \cite{Gupta}, they outline several desirable criteria that a global SA method ought to have. They note that Sobol variance based \cite{Sobol1,Sobol2} and derivative based \cite{Morris} SA only include a subset of these global SA criteria. The desirable global SA criteria outlined in \cite{Gupta} have access to:
 \begin{enumerate}
     \item Local sensitivities (i.e., first-order derivatives),
     \item Global distribution of local sensitivities (characterized, for example, by their mean and variance, or by some other statistics),
     \item Global distribution of model responses (characterized, for example, by their variance, or by some other statistics), and,
     \item Structural organization (shape) of the response surface (including its multi-modality and degree of non-smoothness/roughness).
 \end{enumerate}
The authors later use these to guide the design of the VARS and IVARS \cite{VARS1,VARS2} global SA methods. Further they show how VARS and IVARS can reproduce Sobol and derivative based SA as special cases. The authors discuss how global SA can be used to: assess the similarly of model functions,  find regions of sensitivity, simplify models based on insensitive factors, perform uncertainty apportionment, and identify factor importance, function, and independence. 
 

 VARS performs SA by calculating relevant expectation values of statistics that quantify the variations of a model's response surface. A response surface is defined as the set of outputs of a model, i.e. $\{y\}=\{M(\{v\})\}$, that form a surface in the input space, $\{v\}$, of interest. Typically $y=M(v)$ is a scalar value while $v$ and $\ell$ are vectors in SA, which we will assume is true for the remainder of the article. An important statistic, $S$, is the square difference of the responses $(M(\ell+v)-M(\ell))^2$ between the points $(\ell+v,\ell)$.\footnote{Note that our notation differs from \cite{VARS1, VARS2} in that their $(h,x)$ turns out to be our $(v,\ell)$ due to the differences in interpretation presented in equation (\ref{VARSvarH}).} The VARS expectation values are computed from equally weighted averages of these statistics over the input space,
 $$E(S(M))=\frac{1}{L}\int_{\ell}(M(\ell+v)-M(\ell))^2\,d\ell,$$
having normalization $\frac{1}{L}$. This allows them to construct a multidimensional variogram over $\ell\in L$,
\begin{eqnarray}
\gamma(v)=\frac{1}{2}E((M(\ell+v)-M(\ell))^2)=\frac{1}{2L}\int_{\ell} (M(\ell+v)-M(\ell))^2\,d\ell,\label{vargram}
\end{eqnarray}
which quantifies the expected squared deviation of the response surface at ``scale" $v$. By varying the components of $v$ in the input domain and observing $\gamma(v)$, one learns the expected sensitivity's dependence on scale.  

A key solution presented in \cite{VARS1, VARS2} is to characterize SA across all scales by integrating over them. While preceding methods of global SA have scale dependence (which is seen as a shortcoming of their methods), VARS removes this dependence by calculating the ``integrated variogram",
\begin{eqnarray}
\Gamma(V)=\int_{v\in V}\gamma(v)\,dv,\label{Gamma}
\end{eqnarray}
which is the variogram summed over all scales up in the space of scales $V$, explicitly $\vec{v}\in[0,\vec{V}]$. Analysis with this quantity is called Integrated Variogram Analysis of Response Surfaces (IVARS). In \cite{VARS1, VARS2}, they consider scales $V$ up to the 10\%, 30\% and 50\% of the total input space, which can be used for performing global sensitivity analysis, i.e, for the calculation and investigation of the desirable global SA criteria quantities \cite{Gupta}.

\subsection{IPSA Derivation}
 We will show that IPSA quantifies the probability of a variation of an output due to possible variations stemming from uncertainty in the inputs, for every inquired uncertain input. These distributions are represented in the IPSA probability matrix (\ref{IPSA}), which is the main SA tool in IPSA. From the point of view of current SA methods, IPSA uses measurement uncertainty to weight the probabilistic relevance of each scale $v$ in IVARS. 
Large scale $v$ deviations are suppressed due to their low probability in favor of more probable smaller scale $v$'s near the observed location $\ell$ when the measurement uncertainty is small. 

We make three observations about IVARS that guide the derivation of IPSA: The first observation is that if we divide $\Gamma\rightarrow \Gamma/V$, where $V$ is now a multidimensional volume of the input space, we may interpret equation (\ref{Gamma}) as an expectation value over the variogram,
\begin{eqnarray}
E(\gamma(v))=\int_{v\in V}\frac{1}{V}\gamma(v)\,dv=\frac{\Gamma(V)}{V},\label{GammaH}
\end{eqnarray}
which does not result in a loss of generality. This expectation value is equally written as,
\begin{eqnarray}
E(\gamma(v))=\int_{\ell}\frac{1}{2VL}\Big(\int_{v\in V} (M(\ell+v)-M(\ell))^2\,dv\Big)\,d\ell,\label{VARSvarH}
\end{eqnarray}
where we have switched the order of integration of $\ell$ and $v$. Thus, in the computation of $\Gamma$, one could first integrate over the possible relevant ``scales" $v$ (or in our language ``deviations") differing from the location $\ell$ and only then averaging over all $\ell$, rather than the reverse (as was done in IVARS). 

The second observation is that current SA methods have no stated dependence on the amount of input measurement uncertainty. When considering SA for UP, we believe this is a key missing feature as one would like to know how sensitive their model outcomes are to changes in input measurement uncertainty in practice. To include measurement uncertainty, we begin constructing IPSA by generalizing IVARS to nonuniform input probabilities, $\frac{1}{VL}\rightarrow \rho(v,\ell)$ such that,
\begin{eqnarray}
E(\gamma(v))\longrightarrow\int_{v,\ell}\frac{(M(\ell+v)-M(\ell))^2}{2}\rho(v,\ell)\,dv\,d\ell.
\end{eqnarray}
Reverting back to a uniform probability distribution over $\ell$ and $v$ gives $\Gamma(V)$ from IVARS and letting $\rho(v,\ell)=\delta(v-v')/L$, with $v'$ being the ``actual/single scale", gives back $\gamma(v')$ from VARS as special case statistics of IPSA.\footnote{Later this delta probability distribution would be interpreted as coming from measurement uncertainty, which is a bit unnatural. This shows some amount of negative correlation between the goals of SA from IPSA verses VARS.}

After switching the integration order as we did above, it becomes easier to interpret nonuniform $\rho(v,\ell)\rightarrow\rho(v,\ell|\Delta_{\ell})$ meaningfully as originating from measurement uncertainty. This is done by interpreting $\rho(v|\ell,\Delta_{\ell})$ to be the probability that in reality the value deviates from the observed value of $\ell$ by an amount $v$ due to measurement uncertainty $\Delta_{\ell}$ (for the data-like $x$'s at least). There exist many pdfs that could represent measurement uncertainty in this way, e.g. narrow uniform pdfs centered at the measurement locations; however, we will use Gaussian distributions with $\Delta_{\ell}=\sigma_{\ell}>0$ to represent the inclusion of measurement uncertainty here. That is, a natural choice is,
\begin{eqnarray}
\rho(x|\ell,\Delta_{\ell}=\sigma_{\ell})=\frac{1}{Z}\exp\Big(-\frac{x^2}{2\sigma_{\ell}^2}\Big),\label{gaussh}
\end{eqnarray}
where the observed value $\ell$ plays the role of the measured value that the $x$'s are symmetrically distributed about. One may construct an expectation value that is conditional on a single observation $\ell$, 
\begin{eqnarray}
\Delta^2(\Delta_{\ell},\ell)=\int_{x,\alpha}\frac{(M(\ell+x,\alpha)-M(\ell,\alpha))^2}{2}\rho(x,\alpha|\ell,\Delta_{\ell})\,dx\,d\alpha.\label{IPSAvar}
\end{eqnarray}
Computing this quantity for all $\ell$ allows one to construct $\{\Delta^2(\Delta{\ell},\ell)\}$, which is the set of local expected square deviations of the response surface due to measurement uncertainty $\Delta_{\ell}$. Note that this equation involves the integration over $v$, i.e., over all probabilistically weighted scales at a fixed location $\ell$, whereas in the multidimensional variogram $\gamma(v)$, $\ell$ is integrated over and the scale $v$ is fixed. Thus, it is probabilistically natural to include $\Delta^2(\Delta_{\ell},\ell)$ as an additinoal SA tool in VARS/IVARS as it is conditioned on $\rho(v|\ell)$ whereas $\gamma(v)$ is instead conditioned on $\rho(\ell|v)$ -- both of whose marginalizations over their respective conditioned variable lead to $\Gamma(V)$ in the IVARS limit of IPSA due to the properties of joint probabilities.

The third observation is that expectation values are noninvertable in general and thus they constitute a loss in information. Every expectation value is a many to one map due to the sum. These degenerate results may have distinguishing features that are relevant for SA. 



Although expectation values of linear statistics exhibit undesired positive and negative fluctuation cancellation, i.e. $E[M(v)-E[M(v)]]=0$, if one instead considers the probability of linear statistics, there is no mechanism for cancellation \emph{and} one preserves the entire information content of the model function. That is, by letting,
\begin{eqnarray}
S_{\mbox{\tiny{lin}}}(v,\ell)\equiv S_{\mbox{\tiny{lin}}}=M(\ell+x,\alpha)-M(\ell,\alpha)=\Delta y,
\end{eqnarray}
we have a maximally informative statistic because no information is lost. This statistic informs one about the potential skews in the response, i.e. shorter scale $x$'s may dominate regions for $S_{\mbox{\tiny{lin}}}<0$ as compared to regions of $S_{\mbox{\tiny{lin}}}>0$ or vice versa. We therefore quantify the probabilities of \emph{linear} deviations of the model away from observed values, $p(S_{\mbox{\tiny{lin}}}|\ell,\Delta_{\ell})$, for the purpose of SA. Due to the statistic's linearity, the probabilities for the statistic $S_{\mbox{\tiny{lin}}}$ are simple translation coordinate transformations of the model output probabilities, which we can take advantage of computationally. Probabilistic deviations away from model parameters ($\sim \ell_{\alpha}$) may also be considered in general.

To perform IPSA, we first vectorize the set of relevant probabilities $\{p(y|\ell,\Delta_{\ell},M)\}$. This is, 
\begin{eqnarray}
\mathcal{P}_{y|\Delta_{\ell},\ell}= \mathcal{M}_{y, v}\cdot \mathcal{P}_{v|\Delta_{\ell},\ell}.\label{IPSA}
\end{eqnarray}
We then preform a translation coordinate transformation to quantify deviations from the estimated values $y\rightarrow\Delta y=y-y(\ell)\,\,\forall \ell$ (i.e. for each column of $\mathcal{P}_{y|\Delta_{\ell},\ell}$), which gives,
\begin{eqnarray}
\rightarrow\mathcal{P}_{\Delta y|\Delta_{\ell},\ell}.\label{IPSA}
\end{eqnarray}
This is the IPSA probability matrix.


 In principle, any SA quantity of interest can be inferred or calculated from the IPSA probability matrix, some of them, like expectation values, are easily vectorizable as well. Although a well sampled IPSA probability matrix contains all the information one might be interested in for SA, the result may be too cumbersome to communicate efficiently in its totality -- plotting the set of probabilities in the IPSA probability matrix can only be done in low numbers of dimensions. 

One can use the IPSA probability matrix to calculate coarse grained SA quantities that are bit easier to handle and communicate due to their lower level of detail and dimensionality. 
This might include: confidence intervals that vary with $\ell$, the maximum probability of $y$ per $\ell$, the expected deviation of $y$ per $\ell$, and the variance of $y$ per $\ell$. If one further marginalizes over $\ell$ using $\rho(\ell)$ (i.e. the probability of a measurement at $\ell$), one obtains ``global values", which may be probabilities such as $p(\Delta y|M,\Delta_1,...,\Delta_L)$ or single scalar expectation values. The least informative coarse grained measure is one that summarizes SA with a single scalar value, such as the overall variance of $S_{\tiny{lin.}}$, as there is only so much you can express with one value when the full expression lives in an extremely high dimensional space. For this reason we again stress that the preferred metric for SA is the IPSA probability matrix itself (\ref{IPSA}).

\subsection{IPSA Examples\label{section6_2}}

%

\begin{figure}[H]\centering

\subfloat[]{\label{figure2a}\includegraphics[width=.5\textwidth]{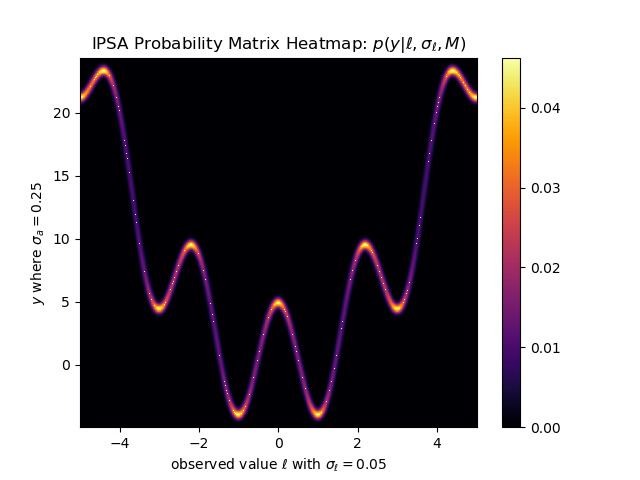}}
\hfill
\subfloat[]{\label{figure2b}\includegraphics[width=.5\textwidth]{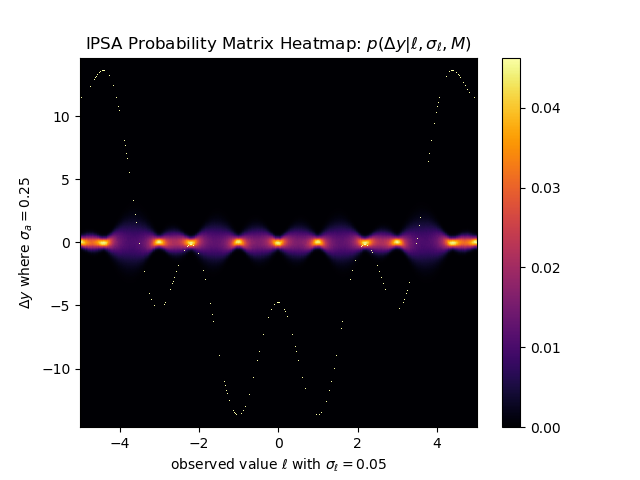}}

\caption{ Low probability (purple) regions are more sensitive to measurement uncertainty because their values are changing more with $\ell$ as can be seen in (a). This can be seen by the larger probability of deviations at these locations in (b). The reverse is true for the high probability (gold) regions. Due to the small measurement uncertainty $\sigma_{\ell}=.05$ relative to the model function, the probability follows the model function curve tightly in (a) and the probability of a deviation away from the estimated value is relatively small in (b).}\label{figure2}
\end{figure}

\begin{figure}[H]\centering

\subfloat[]{\label{figure3a}\includegraphics[width=.5\textwidth]{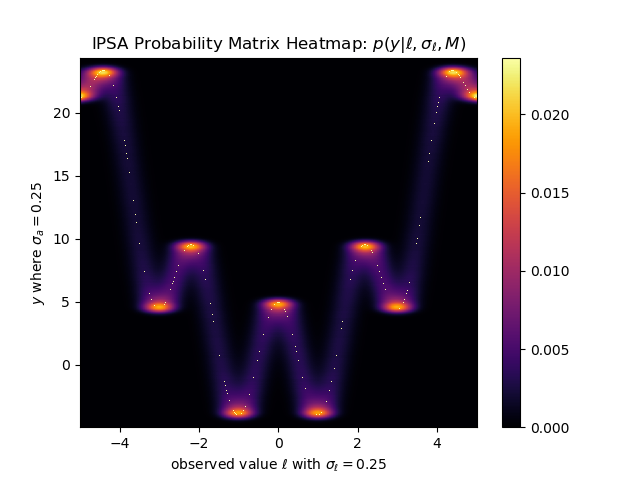}}
\hfill
\subfloat[]{\label{figure3b}\includegraphics[width=.5\textwidth]{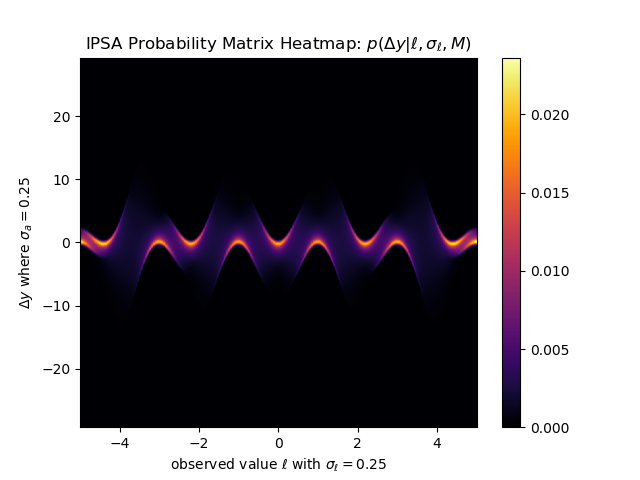}}

\caption{ The measurement uncertainty in $\ell$ has been increased to $\sigma_{\ell}=0.25$ and we begin to see more uncertainty in (a) and larger local asymmetries in the deviations in (b).}\label{figure3}
\end{figure}

\begin{figure}[H]\centering

\subfloat[]{\label{figure4a}\includegraphics[width=.5\textwidth]{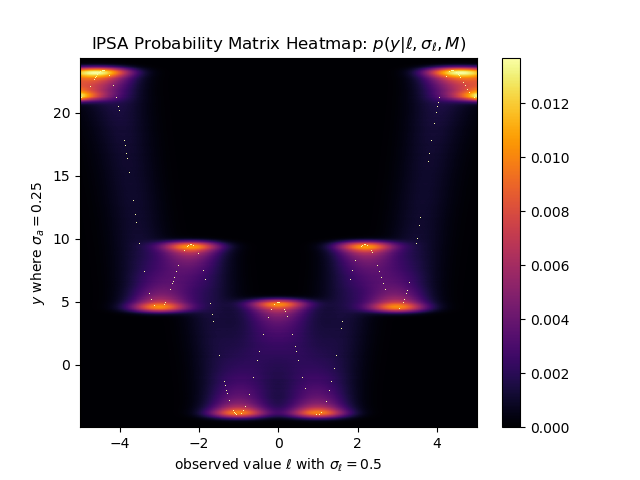}}
\hfill
\subfloat[]{\label{figure4b}\includegraphics[width=.5\textwidth]{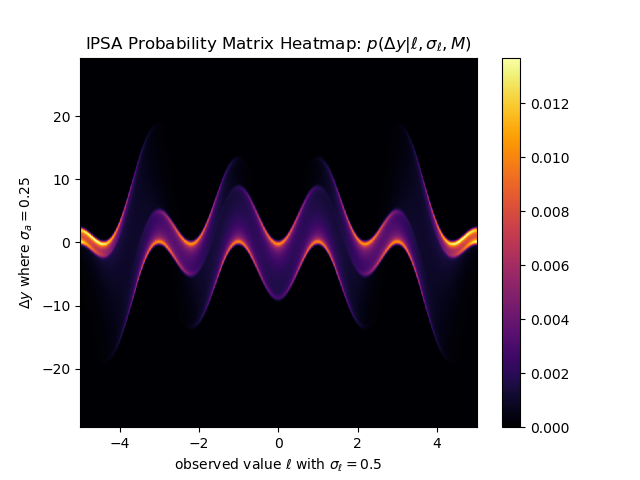}}
\vspace{0mm}
\subfloat[]{\label{figure4c}\includegraphics[width=.5\textwidth]{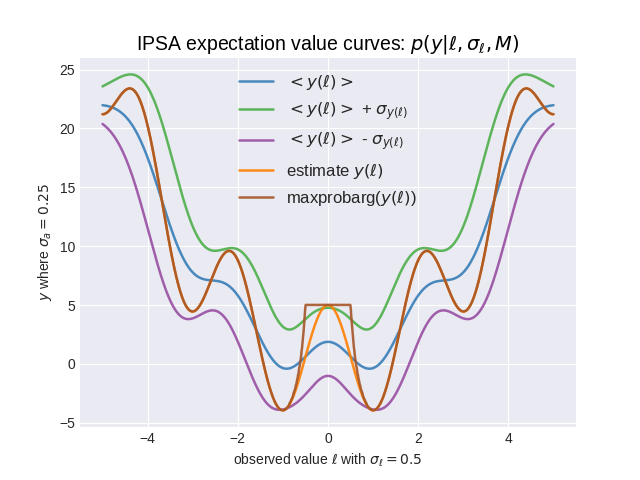}}
\hfill
\subfloat[]{\label{figure4d}\includegraphics[width=.5\textwidth]{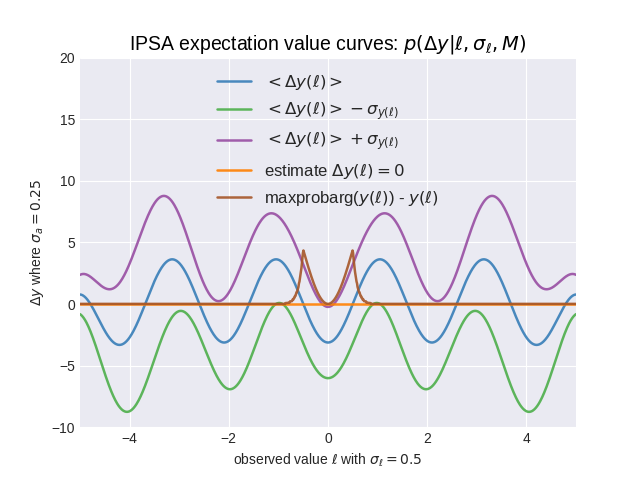}}

\caption{ The measurement uncertainty in $\ell$ has been increased to $\sigma_{\ell}=0.5$ which begins to conflate the probabilities from subsequent waveforms with wavelength $\lambda=\frac{k}{2\pi}\sim0.5$, although there is still a significant dependence on $\ell$ in (a) and (b). The majority of the participating scales are still local around $\ell$ due to their higher probability of occurrence. Figure (c) plots expectation value fields generated from the output probability matrix, the estimated value of $y$,  as well as the values $y$ per $\ell$ with the maximum probability. Plotting the confidence intervals may be useful, although it is not done here. Figure (d) depicts these field lines as linear deviations from the input estimate. Again, the local asymmetries are apparent.
        }\label{figure4}
\end{figure}

\begin{figure}[H]\centering

\subfloat[]{\label{figure5a}\includegraphics[width=.5\textwidth]{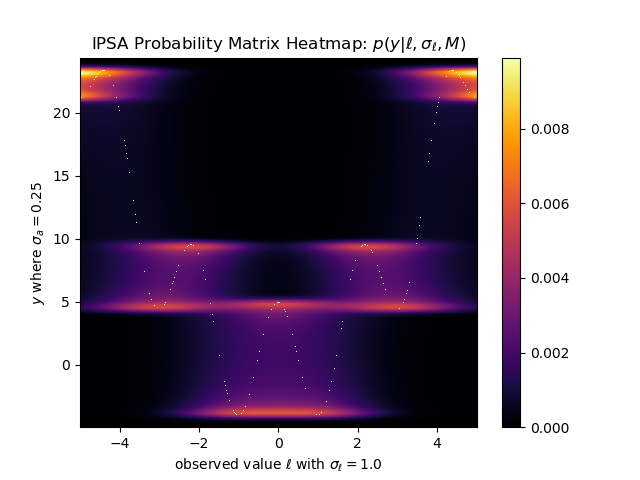}}
\hfill
\subfloat[]{\label{figure5b}\includegraphics[width=.5\textwidth]{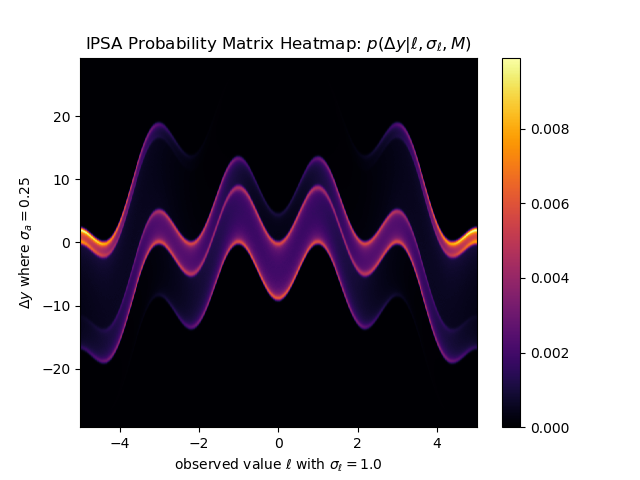}}

\caption{The measurement uncertainty in $\ell$ has been increased to $\sigma_{\ell}=1.0$, which relative to the model function and location, begins to form probability bands across larger deviation scales in (a) and (b).}\label{figure5}
\end{figure}

\begin{figure}[H]\centering

\subfloat[]{\label{figure6a}\includegraphics[width=.5\textwidth]{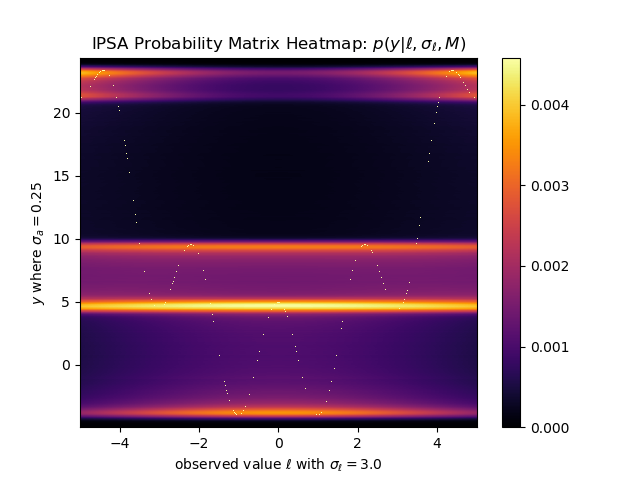}}
\hfill
\subfloat[]{\label{figure6b}\includegraphics[width=.5\textwidth]{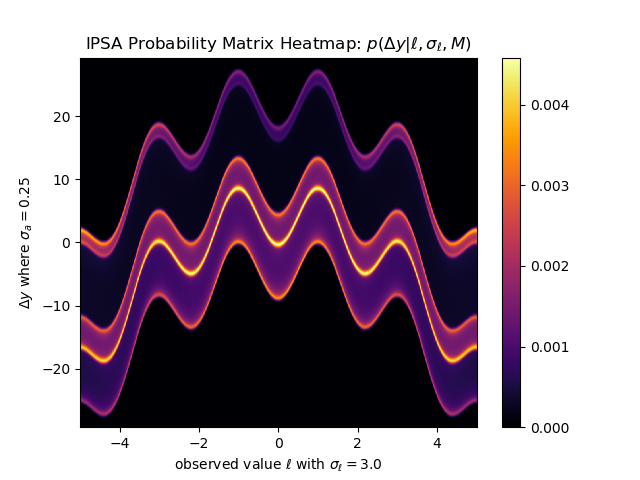}}

\caption{A large measurement standard deviation of $\sigma_{\ell}=3.0$ is causing participation from all scales in (a) and (b) as the input probability is largely uniform, but participation still varies slightly with $\ell$. There is a higher probability of a negative deviation for measurements at locations $|\ell|>3$ in (b) due to the relatively large probability of  $y<10$ across the scales in (a).}\label{figure6}
\end{figure}

In this example we perform IPSA on the following model function,
\begin{eqnarray}
y=M(x,\alpha)= x^2+5\sin(3x)+\alpha,
\end{eqnarray}
which resembles an example function considered in \cite{VARS1,VARS2}. The output probability matrices (a's) and the IPSA probability matrices (b's) are plotted as heat maps in each of Figure \ref{figure2}-\ref{figure6} in correspondence with five Gaussian measurement uncertainty scenarios $\sigma_{\ell}=\{0.05,0.25,0.5,1.0,3.0\}$ (uniform measurement uncertainty across $\ell$). In these scenarios we let $\sigma_{\alpha}=0.25$, which represents the addition of Gaussian noise ($\lambda$) or as an uncertain vertical offset model parameter in $y$ ($\alpha$). Each figure depicts $L=1000$ uncertain measurement locations $\ell\in[-5,5]$ and each distribution is over a $N=10^6$ uniform sampling grid in the space of $(x,\alpha)$. Because we used VUP, the computation times were similar to those simulated in Section \ref{section4}.

The main features to take away from Figures \ref{figure2}-\ref{figure6} are the joint dependence of measurement uncertainty and model functional form that shape of the output probability and IPSA matrices. Of further importance is the amount of relevant information and details that can be inferred from these matrices. We see that increases in input measurement uncertainty cause larger scale deviations to become more probable and, except for very low input uncertainties, that the deviations can be highly asymmetric and location $\ell$ dependent. Knowing that there is a tendency to under or over estimate a value at different locations help better assess the degree of skepticism in the estimated value of an uncertain input.  If higher output accuracy is needed, one should aim to reduce $\sigma_{\ell}$ in the regions where probabilities are more spread out (purple). We see that if measurement uncertainties are large, the high probability regions dominate multiple length scales and measurement location doesn't matter much (Figure \ref{figure6a}), which, if marginalized over $\ell$, is similar to the case one implicitly considers in the large scale case of the IVARS framework.

In detail, the probabilities of the model outputs $p(y|\ell,\sigma_{\ell})$ are plotted in Figures \ref{figure2a}-\ref{figure6a} (one for each $\sigma_{\ell}$, respectively) over $y$ and $\ell$, which depicts the values of the components of the output probability matrix. For each column of pixels (constant $\ell$) in Figures \ref{figure2a}-\ref{figure6a}, VUP discretizes and calculates, 
\begin{eqnarray}
\rho(y|\ell,\sigma_{\ell})&=&\frac{1}{Z}\int_{x,\alpha}\delta(y-M(x,\alpha))\exp\Big(-\frac{(x-\ell)^2}{2\sigma_{\ell}^2}-\frac{\alpha^2}{2\sigma_{\alpha}^2}\Big)\,dx\,d\alpha,
\end{eqnarray}
for each $y$ in the column.\footnote{The coordinate transformation $x\rightarrow x+\ell$ was performed to take the $\ell$ dependence from the model matrix and to put it into the input pdf such that the model matrix could be reused.} The values of $\ell$ correspond to different columns of the output/ IPSA probability matrices. These probabilities are binned into $K=1000$ bins per column for plotting purposes. The $\alpha$'s are considered to have an uncertainty that is independent of the measurement location, which seems to be closer to what happens in most cases, but nothing in the VUP formalism prevents arbitrary amounts of correlation. We truncated the pdf outside the boundary of the input domain as a hard cut off (an impossible region) and renormalized the pdf inside the boundary.


Figures \ref{figure2b}-\ref{figure6b} plot the probability of a deviation $S_{\mbox{\tiny{lin}}}=M(\ell+x)-M(\ell)=\Delta y$ away from the estimated value $y(\ell)\equiv M(\ell,0)=\ell^2+5\sin(3\ell)$, which is the set the arguments where the \emph{input} probability has its maximum values (one could instead consider deviations from the average value(s) if desired), over $y$ and $\ell$. These figures depict the values of the components of the IPSA probability matrix. The maximum input probability value $y(\ell)$ is plotted with a white dotted line in Figures \ref{figure2a}-\ref{figure6a} and in Figure \ref{figure2b} for reference in comparison to Figure \ref{figure2a}. Note the estimated value $y(\ell)$ before the inclusion of measurement uncertainty is not necessarily the same as the $y$ value with the maximum \emph{output} probability after measurement uncertainty is taken into account, as can be seen in Figures \ref{figure4}-\ref{figure6}. 

The function is sampled at a much denser rate than is needed to obtain the same qualitative features (because the function is simple and smooth enough). Instead propagating 100 distinct input probabilities, each with $N=10^4$, takes about 0.005 seconds (+3 seconds if GPU initialization is included) and reveals these same features; however, given access to denser sampling, we used it instead.

\clearpage

 \section{Conclusions}
 
 
In this article we construct VUP, which is a vectorized computational method for efficiently propagating many probability distributions through computationally deterministic model propagators. 
By constructing the model matrix as a probabilistic representation of the model function, and by reusing it for the propagation of $L$ many input probability distributions, VUP has a smaller computational time complexity than MC-type methods for $L\gtrsim 2$. 
If executed on the GPU, and given the matrices are large enough, VUP's computational speed improves further. 

We extrapolate the logic used in VARS and IVARS sensitivity analysis methods to formulate IPSA, which naturally utilizes VUP. The result of IPSA is an informationally dense model output probability matrix that encodes the probability of linear deviations, due to input parametric and measurement uncertainty, on a measurement point by point basis. 
Because IPSA uses VARS and IVARS as its theoretical foundation, it can reproduce their results as special cases as well as the results of Sobol and derivative based SA by extension.  
 

	\paragraph{Acknowledgments}
	This work was supported by the Center for Complex Engineering Systems (CCES) at King Abdulaziz City for Science and Technology (KACST) and the Massachusetts Institute of Technology (MIT). We would like to thank all of the researchers with the Center for Complex Engineering (CCES). Finally, we would also like to thank the members of the Mechatronics Research Lab at MIT, Nicholas Carrara, and Tony Tohme.

\appendix

\section{Computational time complexity\label{appendix}}
We will compare estimates of the computational time complexity of MC and VUP. We revert to the notation $\vec{v}=(\vec{x},\vec{\alpha})$ where $\vec{v}\in\mathbb{R}^n$. The algorithm for MC is:
\begin{algorithm}[H]

\caption{MC - Single probability}
 Returns a probability vector $P(y|M)$ from $N$ samples propagating into $K$ bins:\\\\
$\{\vec{v}_i\} \leftarrow$  Generate $N$ random samples from $\rho(\vec{v}|M)$\\
 $\{y_i\}\leftarrow$ Evaluate the model $\{M(\vec{v}_i)\}$\\
  $\{y_i\}\leftarrow$ Sort($\{y_i\}$)\\
 $P(y|M)\leftarrow$ NormalizeBin($\{y_i\},K$)

\end{algorithm}
To generate a sparse model matrix one saves its set of nonzero elements and their corresponding indices in the matrix. The algorithm for VUP is:

\begin{algorithm}[H]
 Returns a probability vector $P(y|M)$ from $N$ samples propagating into $K$ bins:\\\\
  Generate and collect input/output pairs:\\
\hspace*{6mm}$\{\vec{v}_i\} \leftarrow$ Generate a uniform grid with $N$ samples\\
 \hspace*{6mm}$\{y_i\}\leftarrow$ Evaluate the model $\{M(\vec{v}_i)\}$\\\\
Generate the sparse model matrix $\mathcal{M}_{y,\vec{v}}$:\\
\hspace*{6mm}$\{y_j\},\{[j,i]\}\leftarrow$ Get the UniqueSorted($\{y_i\}$) and its arguments\\
\hspace*{6mm}$\{[j,i]\}\leftarrow$ Are the sparse matrix indices\\
 \hspace*{6mm}$\{\mathcal{M}[j][i]\} \leftarrow$ Set indexed model matrix elements to $1$ (determin. M.M.) \\
\hspace*{6mm}$\mathcal{M}_{y,\vec{v}} \leftarrow \{\mathcal{M}[j][i],[j,i]\}$ is the sparse model matrix \\\\
 $ P(\vec{v}|M) \leftarrow $ Evaluate the input pdf function at $\{\vec{v}_i\}$ and normalize \\
 $ P'(y|M) \leftarrow$ Sparse matrix multiply $\mathcal{M}_{y,\vec{v}} \cdot P(\vec{v}|M) $\\
 $P(y|M)\leftarrow$ MarginalizeBin($P'(y|M),K$)
\caption{VUP - Single probability}

\end{algorithm}

The computational time complexity for MC, for $N>>K>1$ is,
\begin{eqnarray}
C_{MC}\sim \mathcal{O}\Big(N(C_{rn-d}+C_{model})+C_{sort\,bin})\Big),
\end{eqnarray}
where $C_{rn-d}$ is the computational time complexity for the generation of a single $n$-dimensional random number from $\rho(\vec{v}|M)$ and $C_{model}$ is the computational time complexity of evaluating the model at a single $n$-dimensional input. The last piece, $C_{sort\,bin}$, is the estimated time complexity for sorting and binning for $N>>K$, which using mergsort and ordered binning is of order $N\log(N)+N$. This gives 
\begin{eqnarray}
C_{MC}\sim \mathcal{O}\Big(N(C_{rn-d}+C_{model}+\log(N)+1)\Big),
\end{eqnarray}

The computational time complexity for VUP, for $N>>K>1$ is,
\begin{eqnarray}
C_{VUP}\sim\mathcal{O}\Big(N({C}_{pdf}+{C}_{model})+C_{MM}+C_{margbin}+C_{sparse\,mat})\Big),
\end{eqnarray}
where $N*C_{pdf}$ is approximately the computational time complexity for evaluating the pdf $\rho(\vec{v}|M)$ at the $N$ points on the grid and normalizing. We consider the computational complexity for generating the sparse model matrix $C_{MM}$ plus the $C_{margbin}$ to be approximately equal to $C_{sort\,bin}$ from MC. The computational time complexity for the matrix multiplication of an $(S\times N)$ by a $(N\times L)$ matrix is $\mathcal{O}(SNL)$. Here, because we are multiplying by a very sparse matrix with $S\approx 1$ nonzero element in each column (due to normalization) by a single probability vector $(N\times 1)$, the sparse matrix multiplication reduces to $\sim (1\times N)$ by a $(N\times 1)$, which is then order $\mathcal{O}(N)$. Combined this gives, 
\begin{eqnarray}
C_{VUP}\sim\mathcal{O}\Big(N(C_{pdf}+C_{model} + \log(N) +2)\Big).
\end{eqnarray}

Given that the time complexity of generating a $n$-dimensional random number according to $\rho(\vec{v}|M)$ is approximately equal in time complexity to the time complexity to evaluate $\rho(\vec{v}|M)$ a single $n$-dimensional grid point, i.e., given $C_{rn-d}\sim C_{pdf}$, the estimated difference in time complexity of the methods is,
\begin{eqnarray}
C_{MC}-C_{VUP}\sim -\mathcal{O}(N),\label{A5}
\end{eqnarray}
i.e. the result favors MC for the propagation of \emph{a single pdf}. The assumption that $C_{MM}+C_{margbin}\sim C_{sort\,bin}$ and ${C}_{rn-d}\sim C_{pdf}$ is not necessarily needed, but it seems reasonable in theory and was the case in the examples we tried.  

If one wants to propagate $L$ distinct pdfs using MC methods, one must sort, bin, and reevaluate the model for each propagated pdf. Thus, the computational time complexity for this processes using MC is,
\begin{eqnarray}
C_{MC}(L)= L*C_{MC}\sim \mathcal{O}\Big(L*N(C_{rn-d}+C_{model}+\log(N)+1)\Big).
\end{eqnarray}

If we are interested in the propagation of $L$ distinct probability distributions using VUP, and given the model matrix is an accurate representation of our model, we can reuse the model matrix and propagate these probability vectors through sparse matrix multiplication. Thus, the model does not need to be resampled and the values do not need to be resorted or rebinned. One however does have to create $L$ probability distributions to go from probability vectors to matrices $P(\vec{v}|M)\rightarrow \mathcal{P}_{\vec{v},\ell}$ and thus the pdf computation time and the sparse matrix multiplication time increase by a factor of $L$. This means when VUP propagates $L$ probability distributions,
\begin{eqnarray}
C_{VUP}(L)\sim\mathcal{O}\Big(L*N*C_{pdf}+N*C_{model} + N*\log(N)+N +L*N\Big),
\end{eqnarray}
it increases sublinearly in $L$, i.e. $\mathcal{O}\Big(C_{VUP}(L)\Big)<\mathcal{O}\Big(L*C_{VUP}\Big)$ for $L>1$. Thus, the estimated difference in computational time complexity for $L$ propagated pdfs is then,
\begin{eqnarray}
C_{MC}(L)-C_{VUP}(L)\sim\mathcal{O}\Big((L-1)*N*(C_{model}+\log(N)+1)-L*N\Big),\label{A8}
\end{eqnarray}
which is in favor of VUP for $\frac{1}{L-1}\lesssim C_{model}+\log(N)$, which is almost always the case for $L>1$ and from which we can expect large improvements if $C_{model}$ is large. GPU initialization and memory swapping time may be added if the GPU is utilized to perform VUP.

\end{document}